\documentclass{iaus}
\usepackage{graphicx}

\title[The Mons campaign on OB stars] 
{Variability monitoring of OB stars during the Mons campaign}

\author[T. Morel et al.]   
{T. Morel,$^1$ 
 G. Rauw,$^1$ 
 T. Eversberg,$^2$ 
 F. Alves,$^{3}$ 
 W. Arnold,$^{4}$ 
 T. Bergmann,$^{5}$ 
 N. G. Correia Viegas,$^{6}$
 R. Fahed,$^{7}$ 
 A. Fernando,$^{8}$ 
 L. F. Gouveia Carreira,$^{9}$ 
 T. Hunger,$^{10}$ 
 J. H. Knapen,$^{11}$ 
 R. Leadbeater,$^{12}$ 
 F. Marques Dias,$^{13}$ 
 A. F. J. Moffat,$^{7}$ 
 N. Reinecke,$^{14}$ 
 J. Ribeiro,$^{15}$ 
 N. Romeo,$^{16}$ 
 J. S\'anchez Gallego,$^{11}$ 
 E. M. dos Santos,$^{6}$ 
 L. Schanne,$^{17}$ 
 O. Stahl,$^{18}$ 
 Ba. Stober,$^{19}$ 
 Be. Stober,$^{19}$ 
 K. Vollmann,$^{2}$ 
 M. F. Corcoran,$^{20}$ 
 S. M. Dougherty,$^{21}$ 
 K. Hamaguchi,$^{20}$ 
 J. M. Pittard,$^{22}$ 
 A. M. T. Pollock$^{23}$ 
and P. M. Williams$^{24}$}  

\affiliation{
$^{1}$ Institut d'Astrophysique et de G\'eophysique, Universit\'e de Li\`ege, 4000 Li\`ege, Belgium\\
$^{2}$ Schn\"orringen Telescope Science Institute (STScI), Am Kielshof 21a, 51105 K\"oln, Germany\\
$^{3}$ Av. Portugal 616C - 2765--272 Estoril, Portugal\\
$^{4}$ Burggraben 3, 61206 W\"ollstadt, Germany\\ 
$^{5}$ Eichendorffstrasse 8, 63538 Grosskrotzenburg, Germany\\ 
$^{6}$ Rua Nuno Ataide Mascarenhas, No. 47, 2.Esq. 8100--610 Loule, Portugal\\ 
$^{7}$ D\'epartement de Physique, Universit\'e de Montr\'eal, Montr\'eal (Qu\'ebec) H3C 3J7, Canada\\ 
$^{8}$ Alto Ajuda, Rua 27 - No. 215 1300--581, Lisbon, Portugal\\ 
$^{9}$ R. Rego de Agua LT 24 RC Esq., Marrazes, 2400 Leiria, Portugal\\ 
$^{10}$ Normannenweg 39, 59519 M\"ohnesee-K\"orbecke, Germany\\ 
$^{11}$ Instituto de Astrof\'{\i}sica de Canarias, E-38200 La Laguna, Tenerife, Spain\\ 
$^{12}$ Three Hills Observatory, The Birches, Torpenhow CA7 1JF, UK\\ 
$^{13}$ Rua Almirante Campos Rodrigues, Edf. Girassol, 5F, 1500--036 Lisbon, Portugal\\ 
$^{14}$ Fontainegraben 150, 53123, Bonn, Germany\\ 
$^{15}$ R. Venezuela 29 3 Esq. - 1500--618 Lisbon, Portugal\\ 
$^{16}$ Virulylaan 30, 2267 BS Leidschendam, The Netherlands\\ 
$^{17}$ Hohlstrasse 19, 66333 V\"olklingen, Germany\\ 
$^{18}$ ZAH, Landessternwarte K\"onigstuhl, 69117 Heidelberg, Germany\\ 
$^{19}$ Nelkenweg 14, 66907 Glan-M\"unchweiler, Germany\\ 
$^{20}$ Laboratory for High Energy Astrophysics, Goddard Space Flight Center, Greenbelt, USA\\ 
$^{21}$ Herzberg Institute for Astrophysics, Penticton, British Columbia V2A 6J9, Canada \\ 
$^{22}$ School of Physics and Astronomy, The University of Leeds, Leeds LS2 9JT, UK\\ 
$^{23}$ ESA XMM-Newton Science Operations Centre, Villafranca del Castillo, Spain\\ 
$^{24}$ Institute for Astronomy, University of Edinburgh, Royal Observatory, Edinburgh, UK
}

\pubyear{2008}
\volume{xxx}  
\pagerange{119--126}
\jname{Title of your IAU Symposium}
\editors{A.C. Editor, B.D. Editor \& C.E. Editor, eds.}
\begin{document}

\maketitle

\begin{abstract}
We present preliminary results of a 3-month campaign carried out in the framework of the Mons project, where time-resolved H$\alpha$ observations are used to study the wind and circumstellar properties of a number of OB stars.
\keywords{line: profiles, stars: early-type, stars: winds, outflows, stars: individual (HD 14134)}
\end{abstract}

\firstsection 
\section{Context}
The Mons project is a collaboration between professional and amateur astronomers, which was primarily set up to monitor the periastron passage of the colliding-wind binary system WR 140 centred on January 12, 2009 (Fahed et al., this volume).\footnote{See also {\tt http://www.stsci.de/wr140/index\_e.htm}} A dedicated spectroscopic campaign was organised from December 2008 to March 2009 using the 50-cm Mons telescope at Teide Observatory. Time-resolved observations of the H$\alpha$ line (6360--6950 \AA, 0.34 \AA \ pix$^{-1}$) were also obtained for a small sample of early B-type supergiants and Oe stars to investigate the properties of their large-scale wind structures and circumstellar material, respectively. The B1--B3 supergiants were selected from \cite[Morel et al. (2004)]{morel} based on previous indication of cyclical changes (HD 14134 and HD 42087) or strong variations (HD 43384 and HD 52382). Here we present an overview of the variations exhibited by these objects (the data for the Oe stars HD 45314 and HD 60848 are still being reduced) and briefly discuss forthcoming developments in the data analysis.

\section{Preliminary results and perspectives}
\begin{figure}[h]
\begin{center}
\includegraphics[width=5.02in]{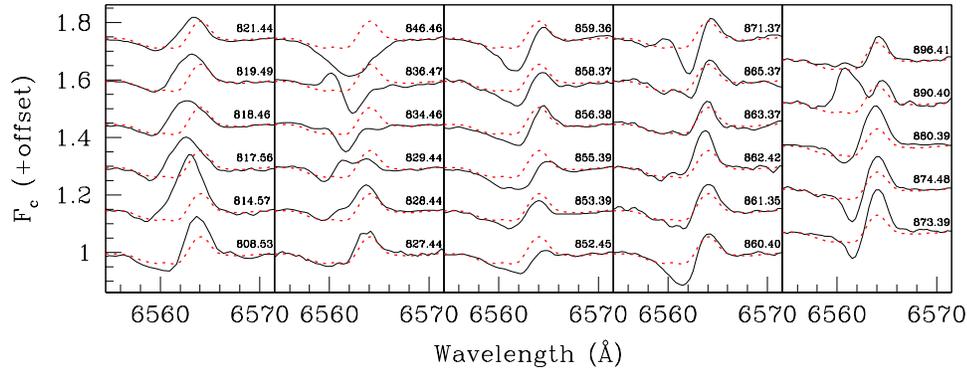} 
\caption{Variations of the H$\alpha$ line in the B3 supergiant HD 14134. The mean profile is overplotted as a dashed line. The date of the observations (HJD--2,454,000) is indicated.}
\label{fig1}
\end{center}
\end{figure}

Variability studies in the UV domain have shown that the winds of OB stars are likely made up of large-scale streams (the `co-rotating interaction regions'; CIRs) whose formation may be triggered by the existence of non-uniform physical conditions at the stellar surface (due, e.g., to magnetic structures or pulsations; \cite[Cranmer \& Owocki 1996]{cranmer}). Optical wind lines can also be used to probe the physical properties of these structures. In particular, revealing rotational modulation in these features would provide evidence that the CIRs extend relatively close to the star and are possibly directly emerging from the photosphere. Strong, daily line-profile variations are observed in all the targets, as illustrated in the case of HD 14134 in Fig.\ref{fig1} by the great variety of profiles observed (strong emission/absorption, double peaked, classical or even inverse P-Cygni profile). This star is of particular interest because of the previous detection of a 12.8-d periodic signal both in photometry and in spectroscopy (\cite[Morel et al. 2004]{morel}).
 
Our efforts will now be directed towards the detection of a periodic behaviour that could allow us to identify the physical processes that drive the variations. For instance, a dipole magnetic field tilted with respect to the rotational axis
 in the Oe stars is expected to induce changes modulated by the rotational period, whereas the variations should take place on much longer timescales if they arise from some kind of disk instability. On the other hand, high-resolution spectroscopic observations of the B3 supergiant HD 14134 are scheduled in November 2010 at OHP (France) to examine the existence of pulsations and to eventually link the variations taking place in the photosphere to those in the wind.


\end{document}